# Charge carrier localization induced by excess Fe in the Fe$_{1+y}$(Te, Se) superconducting system

T.J. Liu[1], X. Ke[2], B. Qian[1], J. Hu[1], D. Fobes[1], E. K. Vehstedt[1], H. Pham[3], J.H. Yang[4], M.H. Fang[1,4], L. Spinu[3], P. Schiffer[2], Y. Liu[2], and Z.Q. Mao[1*]

[1] Department of Physics, Tulane University, New Orleans, Louisiana 70118, USA

[2] Department of Physics and Materials Research Institute, The Pennsylvania State University, University Park, Pennsylvania 16802, USA

[3] Advanced Materials Research Institute and Department of Physics, University of New Orleans, New Orleans, Louisiana 70148, USA

[4] Department of Physics, Zhejiang University, Hangzhou 310027, China

Abstract

We have investigated the effect of Fe nonstoichiometry on properties of the Fe$_{1+y}$(Te, Se) superconductor system by means of resistivity, Hall coefficient, magnetic susceptibility, and specific heat measurements. We find that the excess Fe at interstitial sites of the (Te, Se) layers not only suppresses superconductivity, but also results in a weakly localized electronic state. We argue that these effects originate from the magnetic coupling between the excess Fe and the adjacent Fe square planar sheets, which favors a short-range magnetic order.

**I. INTRODUCTION**

The discovery of high temperature superconductivity in Fe-based compounds has generated tremendous excitement.[1-6] $Fe_{1+y}$(Te, Se) is an important ferrous superconducting system. The superconducting transition temperature of the end member, FeSe ($T_c \approx$ 8K), whose superconductivity was first discovered by Hsu *et al.*,[7] can be raised to 14-15 K by partial Te substitution for Se,[8-9] and up to ~ 27-37 K by applying hydrostatic pressure.[10-13] Density functional theory (DFT) calculations showed that Fermi surfaces (FS) of FeSe and FeTe are similar to those of FeAs compounds,[14] which was confirmed by recent photoemission studies.[15] Superconductivity in $Fe_{1+y}$(Te, Se) is extremely sensitive to stoichiometry,[16] and cannot be understood in the standard electron-phonon picture.[14, 17-18] The observation of spin resonance below $T_c$ and enhancement of spin fluctuations near $T_c$ in this system suggests a superconducting pairing mechanism mediated by spin fluctuations [19-20]. While this material series possesses a crystal structure resembling those of iron arsenides,[7] with Fe square planar sheets (Fe(1) in Fig. 1) forming from the edge-sharing iron chalcogen tetrahedral network, it exhibits an interesting aspect: the interstitial sites of the (Te, Se) layers allow partial occupation of iron, resulting in non-stoichiometric composition $Fe_{1+y}$(Te, Se), where $y$ represents excess Fe at interstitial sites ( Fe(2) in Fig.1).[21-22] This structural characteristic is analogous to that of $Li_{1-x}FeAs$ in which Li occupies interstitial sites of As layers.[23-25]

The end member $Fe_{1+y}Te$ in the $Fe_{1+y}$(Te, Se) series is not superconducting. Instead it exhibits a simultaneous structural and antiferromagnetic (AFM) phase transition near 60-70 K,[21-22, 26] with the AFM structure distinct from those seen in undoped FeAs compounds.[27-29] The AFM order in $Fe_{1+y}Te$ propagates along the diagonal direction of the Fe square lattice,[22, 26] while in FeAs compounds the propagation direction of the SDW-type AFM order is along the edge of the Fe square lattice.[27-29] Another interesting feature of $Fe_{1+y}Te$ is that its AFM wave vector can be tuned by the excess Fe, changing from commensurate to incommensurate when $y$ is increased above 0.076.[22] These results suggest that the mechanism of magnetism in $Fe_{1+y}Te$ should be very different from that of the Fermi surface (FS) nesting driven SDW order in FeAs parent compounds. In fact, FS nesting associated with the SDW instability, as well as the expected SDW gap, was not observed in $Fe_{1+y}Te$.[15, 30] Several theoretical models have been proposed to explain the unusual magnetic order in $Fe_{1+y}Te$.[31-34]

The observation of the dramatic effect of excess Fe on the AFM order in $Fe_{1+y}Te$ suggests that the effect of excess Fe on the superconducting properties of $Fe_{1+y}(Te, Se)$ must be understood in order to understand the superconducting pairing mechanism. In this article, we report experimental results showing how the superconducting and normal-state properties of the $Fe_{1+y}(Te, Se)$ system are controlled by the excess Fe: the excess Fe was found to lead to weakly localized electronic states and the suppression of superconductivity. We argue that the weakly localized states are caused by magnetic coupling between the excess Fe and the adjacent Fe square planar sheets.

**II. EXPERIMENT**

Two groups of single crystals with nominal compositions $Fe_{1+y}Te$ and $Fe_{1+y}(Te_{0.6}Se_{0.4})$, $y = 0$ and 0.18, were used in this study. From our previous studies on polycrystalline samples,[8] the composition $Fe_{1+y}(Te_{0.6}Se_{0.4})$ should have the highest $T_c$, while $Fe_{1+y}Te$ is not superconducting. The single crystals were synthesized by a flux method. Mixed powders of these compositions were sealed in evacuated quartz tubes, and slowly heated to 930 °C, and slowly cooled to 400 °C at a rate of 3 °C/hr before the furnace was shut down. Single crystals with dimensions of ~1 cm × 0.5 cm × 0.5 cm can easily be obtained with this method, and are shown to be the pure *β*-phase (called *α*-phase in Refs. 7-8) with the *P4/nmm* space group by x-ray diffraction. The sample compositions were analyzed using an energy dispersive x-ray spectrometer (EDXS). Table 1 summarizes all nominal and measured compositions of samples used in this study. We will use SC1 and SC2 to denote the superconducting samples with less and more excess Fe, and NSC1 and NSC2 for the non-superconducting parent compounds, respectively. Samples SC1 and NSC1 have approximately 3-4% excess Fe, whereas Samples SC2 and NSC2 have approximately 11% excess Fe. We have performed comprehensive studies on these samples through measurements of resistivity (using a four-probe method), Hall effect (using a five-probe method with magnetic field applied along the *c*-axis), specific heat (a standard semiadiabatic heat pulse technique), and magnetic susceptibility. These measurements were performed in Quantum Design PPMS and SQUID magnetometer.

## III. RESULTS AND DISCUSSIONS

Figure 1a shows the magnetic susceptibility of the samples SC1 and SC2 measured under a magnetic field of 30 Oe with a zero field cooling (ZFC). Sample SC1 was found to show large diamagnetism reflecting bulk superconductivity, consistent with a recent report of bulk superconductivity in single crystal samples $FeTe_{0.5}Se_{0.5}$.[35] Sample SC2, on the other hand, shows small diamagnetism and a broad transition, suggesting the absence of bulk superconductivity. In-plane resistivity ($\rho_{ab}$) measurements showed an onset superconducting transition temperature $T_c^{onset}$ of ~14.8 K for Sample SC1 and 11.6 K for Sample SC2 (Fig. 1b). The transition width $\Delta T_c$, was found to be 1.3 K for Sample SC1 and 5.1 K for Sample SC2. These observations suggest that the excess Fe lowers not only the $T_c$ but the superconducting volume fraction in $Fe_{1+y}(Te_{0.6}Se_{0.4})$ as well. A similar trend was found previously in $Fe_{1+y}Se$.[16]

The physical origin of suppression of superconductivity by the excess Fe in $Fe_{1+y}(Te_{0.6}Se_{0.4})$ can be inferred from its effects on the normal-state transport properties. As shown in Fig. 2a and 2b, the sample SC2 with high excess Fe shows non-metallic behavior above the superconducting transition in both $\rho_{ab}$ and $\rho_c$, with logarithmic temperature dependences below 50 K (Fig. 3a). The logarithmic behavior is characteristic of weak localization in two dimensions. Sample SC1 containing less excess Fe, on the other hand, features metallic behavior below 200 K for $\rho_{ab}$ (Fig. 2a) and below 75 K for $\rho_c$ (Fig. 2b). These observations suggest that the increase of excess Fe concentration results in a weakly localized electronic state.

The behavior of the non-superconducting parent compounds $Fe_{1+y}Te$ is also strongly affected by the excess Fe. As seen in Fig. 4, the sample NSC1 with low excess Fe exhibits metallic behavior below the structural and AFM transitions at $T_{ST}$ = 72 K in both $\rho_{ab}$ and $\rho_c$, consistent with early results on our polycrystalline samples.[8] Moreover, quadratic temperature dependence below 28 K was observed in this sample, as shown in Fig. 3b. This feature, together with the constant electronic specific coefficient observed at low temperatures (see below), suggests a Fermi liquid (FL) ground state. However, for the sample NSC2 with more excess Fe and $T_{ST}$ = 65 K, both $\rho_{ab}$ and $\rho_c$ show weakly non-metallic behavior following a minimum below $T_{ST}$, suggesting a weakly localized electronic state, similar to the phenomena seen in the sample SC2. We note that both SC2 and NSC2 samples exhibit a kink in $\rho_c$ near 120K; this feature seems associated with the ~125 K magnetic anomaly observed previously in polycrystalline samples.[8] Its origin has not yet been clarified.

The influence of excess Fe on electronic states is further revealed in the temperature dependences of Hall coefficient $R_H$ of these samples, as shown in Fig. 5. $R_H$ is determined from the slope of Hall resistivity $\rho_{xy}$ ($H$) (Inset of Fig. 5a). Both SC1 and SC2 have positive $R_H$ above $T_c$, indicating that their electronic transport is dominated by holes. The temperature dependence of $R_H$, however, revealed significant differences between them: sample SC1 exhibits a broad peak at approximately 60 K, while sample SC2 exhibits an upturn below 50 K, below which logarithmic temperature dependence was seen in resistivity. The increasing $R_H$ at low temperatures seen in SC2 implies that the charge carry density decreases with decreasing temperature, consistent with the localization behavior seen in the temperature dependence of resistivity. The broad peak in sample SC1 is similar to that seen in the Co-doped BaFe$_2$As$_2$ superconductor.[36] Samples NSC1 and NSC2 were also found to display marked differences in the temperature dependence of $R_H$. Sample NSC1 has positive $R_H$ at temperatures above $T_{ST}$, and shows a sharp drop near $T_{ST}$, from ~ 1.5×10$^{-9}$ m$^3$/C to a negative value of ~ -2×10$^{-9}$ m$^3$/C, consistent with a previous result obtained on a similar sample.[30] In contrast, $R_H$ for Sample NSC2 was found to exhibit a slight decrease below $T_{ST}$, remaining positive for the whole temperature range measured. These observations indicate that the increase in excess Fe results in a significant change in carrier density in both superconducting and non-superconducting samples.

Specific heat data taken on these samples show a significant anomaly at $T_c$ for SC1 and no discernable feature for SC2 (Fig. 6), confirming the absence of bulk superconductivity in the latter as suggested above. For samples NSC1 and NSC2, the low-temperature specific heat data can be well described by $C = \gamma T + \beta T^3 + \delta T^5$, where $\gamma T$ and $\beta T^3 + \delta T^5$ are electron and phonon specific heat respectively. Plotting $C/T$ against $T^2$ and fits to experimental data below 12 K (Inset to Fig. 6b) yield an estimate of $\gamma \approx 33$ mJ/mol K$^2$ for NSC1 and 27 mJ/mol K$^2$ for NSC2. The decrease in $\gamma$ caused by the increase of excess Fe concentration suggests that the excess Fe gives rise to a decrease of the density of states (DOS) near the Fermi level, consistent with the differences in resistivity and Hall coefficient between samples with more and less excess Fe.

The charge carrier localization observed in Fe$_{1+y}$(Te,Se) system with excess Fe can be attributed to the magnetic coupling between the excess Fe and the adjacent Fe square planar sheets. DFT calculations on Fe$_{1+y}$Te showed that the excess Fe at interstitial sites of (Te,Se) layers is magnetic,[37] in good agreement with the neutron scattering experiment showing that the excess Fe at interstitial sites (Fe(2)) and the Fe on the square lattice (Fe(1)) are magnetically coupled, following the same magnetic modulation.[22] This coupling plays an important role in tuning the AFM order from commensurate to incommensurate when the excess Fe content is increased.[22] Furthermore, DFT calculations[37] indicate that in Fe$_{1+y}$Te the electronic states near $E_F$ are dominated by the Fe(1) 3$d$ characteristics with smaller contributions from the Fe(2) 3$d$ band; the formation of local moments at Fe(2) sites not only reduces the Fe(2) 3$d$ DOS at $E_F$, but also considerably decreases the Fe(1) 3$d$ DOS at $E_F$, resulting in a pseudogap near $E_F$. These theoretical results are consistent with our experimental results. Therefore the Fe(1)-Fe(2) magnetic coupling is the driving force for the charge carrier localization in Fe$_{1+y}$Te.

The magnetic coupling between Fe(1) and Fe(2) is also responsible for the lack of bulk superconductivity and the charge carrier localization observed in SC2. While this sample does not exhibit a long-range magnetic order, a short-range magnetic order was observed in both polycrystalline[22] and single-crystal[38] samples with the excess Fe content close to that in SC2. In addition, the magnetic susceptibility of SC2 is ~0.02 emu/mol right above $T_c$, which is about one order of magnitude larger than that of SC1. These observations suggest that Fe(2) is magnetic in SC2 and that its magnetic coupling with Fe(1) is essential for facilitating the short-range magnetic order. Moreover, our previous neutron scattering measurements revealed that the short-range magnetic order in samples with rich excess Fe enhances noticeably below 40 K.[22] This characteristic temperature is close to the temperature below which the charge carrier localization behavior in both resistivity (Fig. 2-3) and Hall coefficient (Fig. 5a) was observed, suggesting that the short-range magnetic order stabilized by the Fe(1)-Fe(2) magnetic coupling is the origin of the charge carrier localization in the superconducting sample containing high excess Fe. The small superconducting volume fraction observed in such samples suggests that the Fe(1)-Fe(2) magnetic coupling that mediates the short-range magnetic order suppresses superconducting pairing interaction. Disorders/defects introduced by the presence of excess Fe or Se substitution for Te should only have a minor effect on superconductivity since the Fe$_{1+y}$(Te,Se) system is believed to have a considerably high upper critical field and a short coherence length as the LaFeAs(O,F) and $(Ba,K)Fe_2As_2$ systems.[30, 39-40]

In general, short-range magnetic correlation could lead to an anomaly in specific heat at low temperatures. In sample SC2 with rich excess Fe we did not observe any feature related to short-range magnetic correlations. This can be attributed to the fact that this sample, while it does not exhibit bulk superconductivity, undergoes an inhomogeneous, non-bulk superconducting transition, as shown in the susceptibility data in Fig. 1a; this would decrease electronic specific heat, thus smearing the anomaly feature caused by short-range magnetic correlations.

## IV. CONCLUSIONS

In summary, we have investigated the role of excess Fe at interstitial sites of (Te,Se) layers in the $Fe_{1+y}$(Te,Se) superconductor system and shown that in the superconducting $Fe_{1+y}$(Te,Se) the excess Fe not only suppresses superconductivity but also leads to weak charge carrier localization; in the non-superconducting parent compound $Fe_{1+y}Te$, however, the increase of excess Fe concentration causes an electronic state evolution from a FL to a weakly localized state. Together with the magnetic structure previously established by neutron scattering studies and recent DFT calculations, our results suggest that such weak charge carrier localization is related to the magnetic coupling between the excess Fe and the adjacent Fe sheets, which is responsible for the superconductivity suppression caused by the excess Fe.

## ACKNOWLEDGMENTS

The authors thank Dr. W. Bao for useful discussions. The work at Tulane is supported by the NSF under grant DMR-0645305 for materials and equipment, the DOE under DE-FG02-07ER46358 for personnel, the DOD ARO under W911NF-08-C-0131, and the Research Corporation. Work at UNO is supported by DARPA through Grant No. HR0011-07-1-0031. Work at Penn State is supported by the DOE under Grant DE-FG02-04ER46159 and DOD ARO under Grant W911NF-07-1-0182 and NSF under Grant DMR-0701582. Work at Zhejiang University is supported by the NBRPC (No.2006CB01003, 2009CB929104) and the PCSIRT of the Ministry of Education of China (IRT0754).

∗ Corresponding author: zmao@tulane.edu

Table 1	Sample nominal compositions and compositions measured by EDXS.

| Nominal composition | Measured average composition | Label |
|---|---|---|
| Fe(Te$_{0.6}$Se$_{0.4}$) | Fe$_{1.03}$(Te$_{0.63}$Se$_{0.37}$) | SC1 |
| Fe$_{1.18}$(Te$_{0.6}$Se$_{0.4}$) | Fe$_{1.11}$(Te$_{0.64}$Se$_{0.36}$) | SC2 |
| FeTe | Fe$_{1.04}$Te | NSC1 |
| Fe$_{1.18}$Te | Fe$_{1.11}$Te | NSC2 |

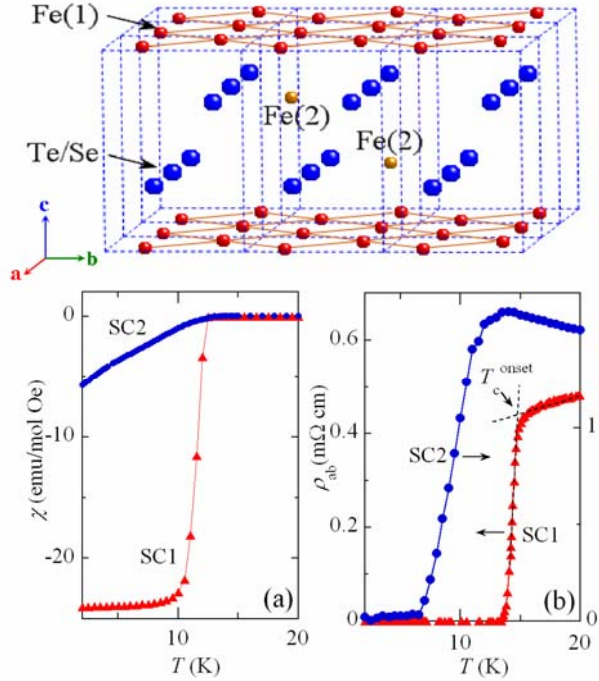

Figure 1: Upper panel: Schematic crystal structure of $Fe_{1+y}(Te, Se)$. The iron on the square planar sheet is denoted by Fe(1); the iron partially occupying at the interstitial sites of the (Te, Se) layers is the excess Fe, denoted by Fe(2). (a) Magnetic susceptibility as a function of temperature $\chi(T)$ measured under a magnetic field of 30 Oe (applied along the $c$-axis). (b) In-plane resistivity as a function of temperatures $\rho_{ab}(T)$. SC1 and SC2 represent two superconducting samples with 3% and 11% Fe(2).

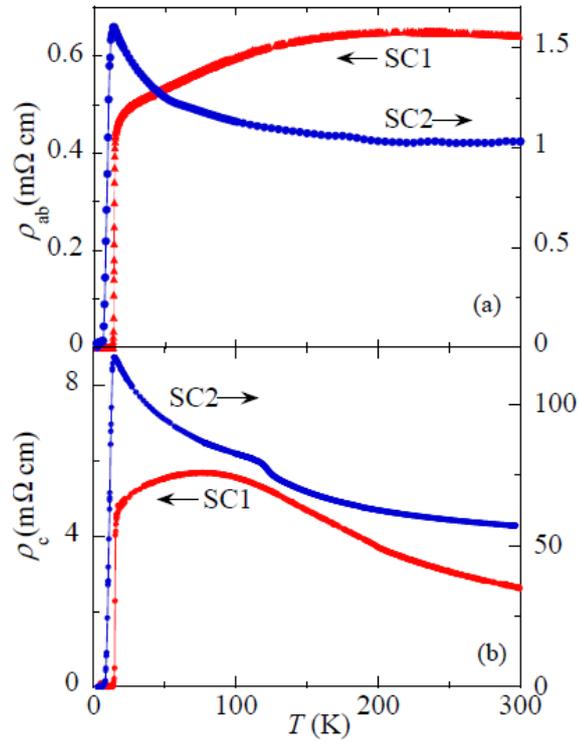

Figure 2: (a) In-plane resistivity as a function of temperature $\rho_{ab}(T)$ for superconducting samples SC1 (3% Fe(2)) and SC2 (11%Fe(2)) . (b) Resistivity along the *c*-axis $\rho_c(T)$ for samples SC1 and SC2.

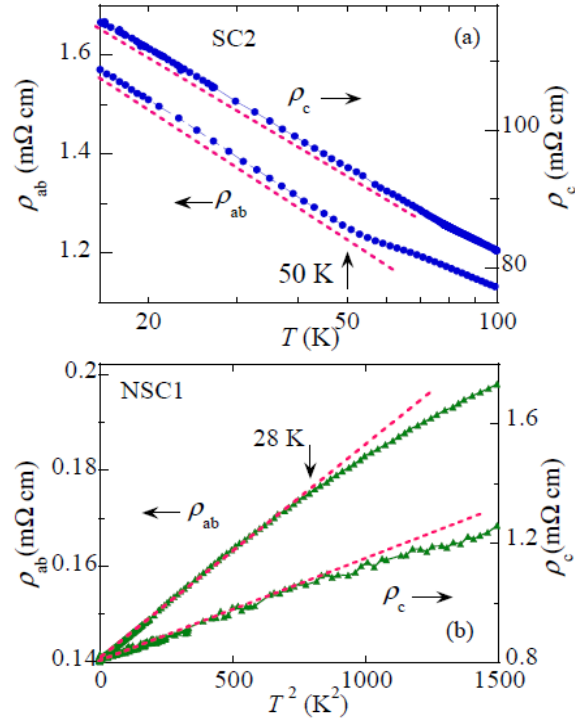

Figure 3: (a) Resistivity as a function of temperature plotted on the $\log T$ scale for superconducting sample SC2(11%Fe(2)). (b) ) Resistivity as a function of temperature plotted on the $T^2$ scale for non-superconducting sample NSC1 (4% Fe(2)) .

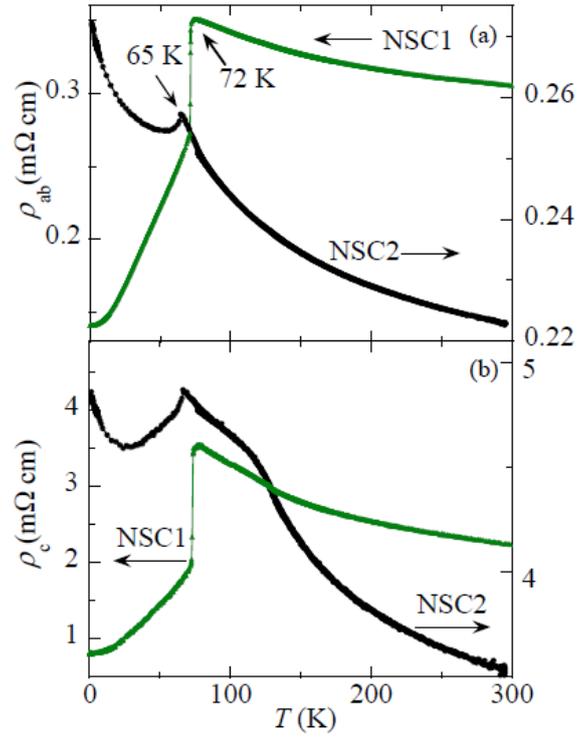

Figure 4: (a) In-plane resistivity as a function of temperature $\rho_{ab}(T)$ for non-superconducting samples NSC1 (4% Fe(2)) and NSC2 (11% Fe(2)). (b) Resistivity along the $c$-axis as a function of temperature $\rho_c(T)$ for NSC1 and NSC2.

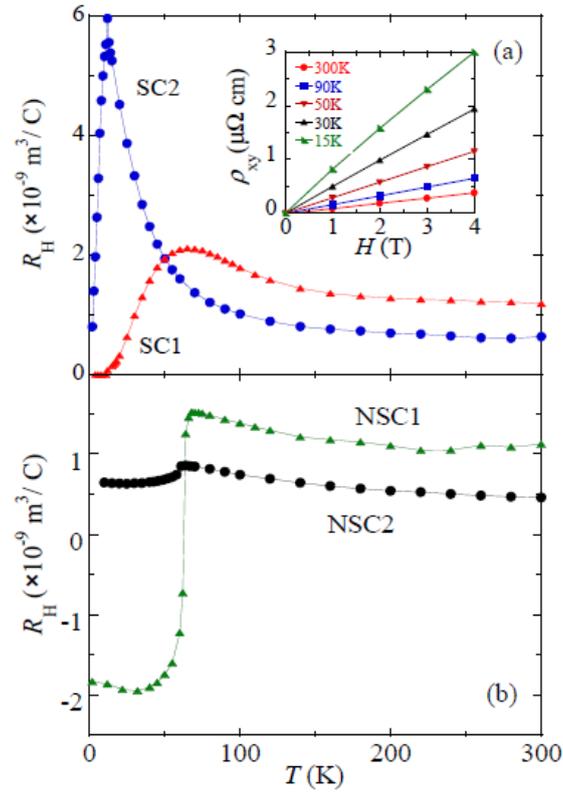

Figure 5: (a) Hall coefficient $R_H$ as a function of temperature for superconducting samples SC1 (3% Fe(2)) and SC2 (11%Fe(2)). (b) Hall coefficient $R_H$ as a function of temperature for non-superconducting samples NSC1 (4% Fe(2)) and NSC2 (11% Fe(2)). Inset in (a): Hall resistivity $\rho_{xy}$ vs. magnetic field at various temperatures for sample SC2.

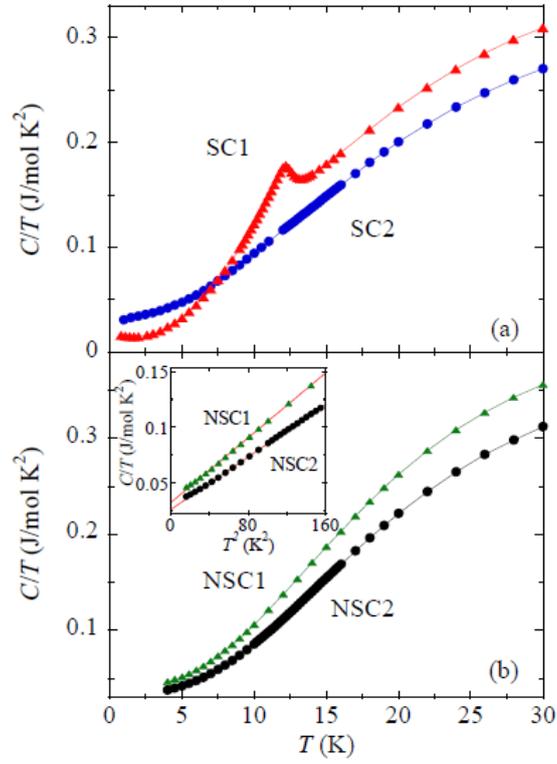

Figure 6: Heat capacity $C/T$ as a function of temperature for superconducting samples SC1, SC2 (a), and non-superconducting samples NSC1, NSC2 (b). Inset in (b) shows $C/T$ versus $T^2$. The solid lines represent the fit to experimental data below 12 K (see the text).